# 5G NR-V2X: Towards Connected and Cooperative Autonomous Driving

Hamidreza Bagheri, Md Noor-A-Rahim, Zilong Liu, Haeyoung Lee, Dirk Pesch, Klaus Moessner, and Pei Xiao

*Abstract*—This paper is concerned with the key features and fundamental technology components for 5G New Radio (NR) for genuine realization of connected and cooperative autonomous driving. We discuss the major functionalities of physical layer, Sidelink features and its resource allocation, architecture flexibility, security and privacy mechanisms, and precise positioning techniques with an evolution path from existing cellular vehicle-to-everything (V2X) technology towards NR-V2X. Moreover, we envisage and highlight the potential of machine learning for further enhancement of various NR-V2X services. Lastly, we show how 5G NR can be configured to support advanced V2X use cases in autonomous driving.

## I. INTRODUCTION

The fifth generation (5G) mobile communication networks, aiming for highly scalable, converged, and ubiquitous connectivity, will be a game changer in opening the door to new opportunities, services, applications, and a wide range of use cases. One of the most promising 5G use cases, expected to shape and revolutionize future transportation, is vehicle-to-everything (V2X) communication, which is seen as a key enabler for connected and autonomous driving. V2X communications, as defined by the 3rd generation partnership project (3GPP), consists four types of connectivity: vehicle-to-vehicle (V2V), vehicle-to-pedestrian (V2P), vehicle-to-infrastructure (V2I), and vehicle-to-network (V2N).

Next generation vehicles will be equipped with cameras, radar, global navigation satellite system (GNSS), wireless technologies, and various types of sensors to support autonomous driving at different levels. However, the functionality of these embedded sensors and cameras is limited by the need for line-of-sight propagation. This can be circumvented by equipping vehicles with cellular V2X (C-V2X) technology complementing embedded sensor functions by sensor data exchange between vehicles, thus providing a higher level of driver situational awareness. So far, C-V2X has attracted significant interests from both the academic and industrial communities. A very promising technology to realize V2X communications and autonomous driving is 5G New Radio (NR). The 5G network is expected to provide ultra-high reliability, low-latency, high throughput, flexible mobility, and energy efficiency. From a communication point of view, 5G should support the following three broad categories of services: enhanced mobile broadband (eMBB), massive machine-type communications (mMTC) and ultra-reliable low-latency communications (URLLC). Specifically, eMBB, aiming to provide data rates of at least 10 Gbps for uplink and 20 Gbps for downlink channels, plays a pivotal role for in-car video conferencing/gaming, various multimedia services, or high-precision map downloading, etc. mMTC will allow future driverless vehicles to constantly sense and learn environment changes from embedded sensors deployed in cars or within the infrastructure. URLLC targets 1 ms over-the-air round trip time (RTT) and 99.999% reliability for single transmissions, which are critical for autonomous driving.

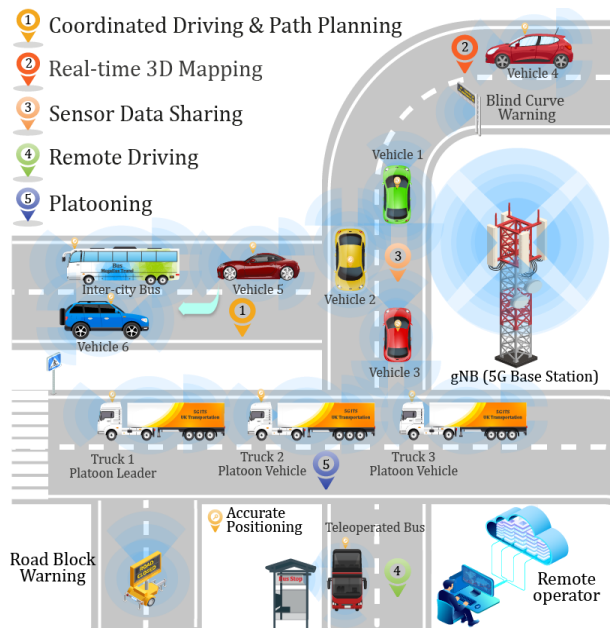

Fig. 1. Advanced use cases and services of 5G NR-V2X.

In fact, 5G NR is designed as a unified framework to address a wide range of the service requirements and to enable novel V2X use cases, as illustrated in Fig. 1.

H. Bagheri, H. Lee, and P. Xiao are with the University of Surrey. (e-mail:{hamidreza.bagheri, haeyoung.lee, pei.xiao}@surrey.ac.uk).

Z. Liu is with the University of Essex, UK. (zilong.liu@essex.ac.uk)

K. Moessner is with the University of Surrey and Chemnitz University of Technology, Chemnitz, Germany. (email: klaus.moessner@etit.tu-chemnitz.de).

M. Noor-A-Rahim and D. Pesch are with the University College Cork, Ireland. (email:{m.rahim, d.pesch}@cs.ucc.ie)

The primary objective of this paper is to indicate how 5G NR supports the realization of autonomous driving. We first discuss the 3GPP standardization roadmap focusing on key features of NR-V2X communications. Next, we present the design considerations, technology components, functionalities, and key enhancements of NR-V2X. We provide an insight into novel and powerful attributes of the NR physical layer (PHY). We discuss NR Sidelink resource allocation and highlight its enhanced functionalities for broadcasting and multicasting. We briefly outline 5G NR architecture deployment options focusing on flexible mobility and dual connectivity. We explain security and privacy issues of NR-V2X communications. In addition, we discuss how machine learning can be exploited to improve the performance of V2X communications. Finally, we discuss advanced use cases for tangible applications of NR-V2X.

## II. 5G NR: 3GPP ROADMAP

The first C-V2X specifications incorporating long-term evolution (LTE) communication technology into vehicular networks, denoted LTE-V2X, were introduced in 3GPP Rel. 14 [1]. In LTE-V2X two operation modes are defined: 1) Network-based communication uses conventional LTE infrastructure to enable vehicles to communicate with the network. The LTE-Uu interface refers to the logical interface between a vehicle and network infrastructure. 2) Direct communication mode is based on device-to-device (D2D) communication, defined in 3GPP Rel. 13 [2]. This mode allows devices to exchange real-time information directly without involving network infrastructure. The PC5 interface, known as LTE Sidelink, is designed to enable direct short-range communications between devices (e.g., V2V, V2P, V2I).

3GPP Rel. 14, completed in June 2017, forms the basis and roadmap for LTE-V2X towards further enhancements and integration into 5G NR. Despite all its capabilities, LTE-V2X does not address the stringent requirements for autonomous driving, specifically for URLLC. 3GPP has defined a new end-to-end network architecture for 5G NR that can meet the needs for autonomous driving. 3GPP timeline for the 5G standard follows two consecutive phases. Phase 1, Rel. 15, the first step on the 5G NR standardization roadmap, focused on eMBB and initial URLLC [3]. Phase 2 started with 3GPP Rel. 16, and focuses on expanding and optimizing the features developed in Phase 1. NR-V2X in Rel. 16 aims at bringing enhanced URLLC and higher throughput, while maintaining backward compatibility with Rel. 15 [4]. NR-V2X, in addition to broadcast transmissions, will support both unicast and multicast transmissions. 3GPP Rel. 17 will continue to improve 5G coverage, mobility, deployment, latency, and services. Fig. 2 illustrates the 3GPP roadmap towards NR-V2X.

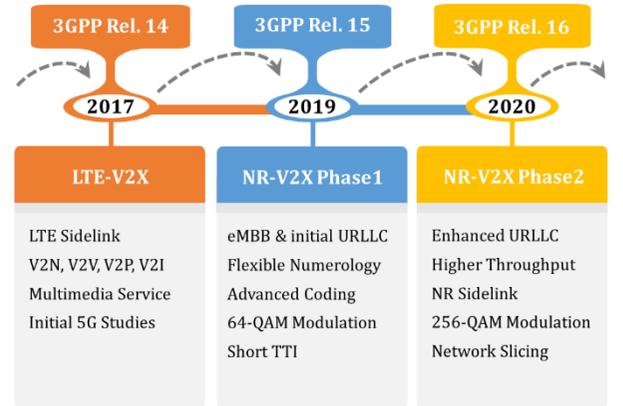

Fig. 2. 3GPP roadmap towards 5G NR-V2X.

## III. KEY FEATURES OF 5G NEW RADIO

This section summarizes NR key features that will fulfill the diverse and stringent requirements of autonomous driving from networks, users and applications perspectives.

### A. The 5G NR Physical Layer (PHY) Design

The 5G NR PHY design needs to deal with harsh V2X channel conditions and diverse data service requirements, specifically: 1) Highly dynamic mobility from low-speed vehicles (e.g., less than 60 km/h) to high-speed cars/trains (e.g., 500 km/h or higher). The air interface design for high mobility communication requires more time-frequency resources to deal with the impairments incurred by Doppler spread and multi-path channels. 2) Wide range of data services (e.g., in-car multimedia entertainment, video conferencing, high-precision map downloading, etc) with different quality-of-service (QoS) requirements in terms of reliability, latency, and data rates. Some requirements (e.g., high data throughput against ultra-reliability) may be conflicting and hence it may be difficult to support them simultaneously.

Against this background, the frame structure of 5G NR [5] allows flexible configurations for enabling the support of a majority of C-V2X use cases. Similar to LTE, 5G NR uses orthogonal frequency-division multiplexing (OFDM) whose performance is sensitive to inter-carrier interference (ICI) incurred by carrier frequency offsets and Doppler spreads/shifts. The maximum channel bandwidth per NR Carrier is 400 MHz compared to 20 MHz in LTE. Identical to LTE, the frame length is fixed to 10 ms, the length of a subframe is 1 ms, the number of subcarriers per resource block (RB) is 12, and each slot comprises 14 OFDM symbols (12 symbols for extended cyclic-prefix mode).

Compared to the LTE numerology with subcarrier spacing of 15 kHz, the NR frame structure supports multiple subcarrier spacings including 15, 30, 60, 120, or 240 kHz. A small subcarrier spacing could be configured for C-V2X use cases requiring high data rates but with low/modest mobility, while a large subcarrier spacing is of particular interest for the suppression of ICI in high mobility channels.

Channel coding plays a fundamental role in C-V2X PHY to accommodate a diverse range of requirements in terms of data throughput, packet length, decoding latency, mobility, rate compatibility, and capability of supporting efficient hybrid automatic repeat request (HARQ). Unlike LTE, which uses convolutional and Turbo codes, two capacity-approaching channel codes have been adopted in 5G NR [6]: low-density parity-check (LDPC) codes and polar codes. While the former is used to protect user data, the latter is for control channels in eMBB and URLLC which require ultra-low decoding latency. Excellent quasi-cyclic LDPC (QC-LDPC) codes have been designed for 5G NR. The unique structure of QC-LDPC allows parallel decoding in the hardware implementation (i.e., lower decoding latency).

C-V2X services in 5G NR are expected to share and compete with other vertical applications for system resources (e.g., spectrum/network bandwidth, storage and computing, etc.) within a common physical infrastructure. A central question is how to design an efficient network to provide guaranteed QoS for V2X while balancing data services to other vertical applications. Network slicing (NS), the paradigm to create multiple logical networks tailored to different types of data services and business operators [7], offers a mechanism to meet the requirements of all use cases and enables individual design, deployment, customization, and optimization of different network slices on a common infrastructure. Although initially proposed for the partition of core networks using techniques such as network function virtualization (NFV) and software-defined networking (SDN), the concept of NS has been extended to provide efficient end-to-end data services by slicing PHY resources in radio access networks (RANs). The slicing of PHY resources mainly involves the dynamic allocation of time and frequency resources by providing multiple numerologies, each of which constitutes a set of data frame parameters such as multi-carrier waveforms, sub-carrier spacings, sampling rates, frame and symbol durations.

*B.    NR Sidelink Features and Resource Allocation*

Through Sidelink protocols, each vehicle can directly exchange its own status information, such as location, speed, trajectory and intended local route, with other vehicles, pedestrians, and road infrastructure. The basic functionalities of the NR Sidelink are the same as those in the LTE Sidelink. However, NR Sidelink introduces major enhancements in functionality that enable advanced 5G use cases and could enhance autonomous driving. The key enhancements in the NR Sidelink protocols are as follows: i) Sidelink feedback channel for higher reliability and lower latency, ii) carrier aggregation with support for up to 16 carriers, iii) modulation scheme supporting up to 256-QAM for increased throughput per single carrier, and iv) modified resource scheduling for reduced resource selection time. Moreover, NR-V2X, along with traditional broadcast communication, supports unicast and groupcast communications, where one vehicle can transmit different types of messages with different QoS requirements. For instance, a vehicle can transmit some periodic messages by broadcasting and aperiodic messages through unicast or groupcast. The reliability of unicast and groupcast communications can be improved via a re-transmission mechanism. It is noted that the re-transmission in LTE-V2X is carried out in a blind manner, i.e., when the source vehicle uses re-transmissions, it re-transmits regardless whether the initial transmission was successful or not. In the case of successful transmission however, such blind re-transmission leads to resource wastage. When several transmissions are required, blind re-transmission may be highly inefficient. In NR-V2X, a new feedback channel, called physical sidelink feedback channel (PSFCH), is introduced to enable feedback-based re-transmission and channel state information acquisition [8]. Detailed operations and procedures of PSFCH feedback transmissions is presented in [9].

In NR-V2X, the available resources for direct communication between vehicles can be either dedicated or shared by cellular users. To manage the resources, two Sidelink modes are defined for NR-V2X, Mode-1 and Mode-2. In Sidelink Mode-1, it is assumed that the vehicles are fully covered by one or more base stations (BSs). The BSs allocate resources to vehicles based on configured and dynamic scheduling. Configured scheduling adopts a pre-defined bitmap-based resource allocation, while dynamic scheduling allocates or reallocates resources every millisecond based on the varying channel conditions. In Sidelink Mode-2, resources need to be allocated in a distributed manner without cellular coverage. There are four sub-modes, sub-mode 2(a)-2(d) for Mode-2. In 2(a), each vehicle can select its resources autonomously through a sensing based semi-persistent transmission mechanism. 2(b) is a cooperative distributed scheduling approach, where vehicles can assist each other in determining the most suitable transmission resources. In 2(c), a vehicle selects the resources based on preconfigured scheduling. In 2(d), a vehicle schedules the Sidelink transmissions for its neighbouring vehicles.

### C. Dual Connectivity and Mobility Robustness

3GPP has defined multiple options for 5G NR deployment, which can be broadly categorized into two modes, non-standalone (NSA) and standalone (SA). In order to accelerate the deployment of 5G networks, the initial phase of NR will be aided by existing 4G infrastructure and deployed in NSA operation mode. In contrast, the full version of NR will be implemented and deployed in SA mode.

The NSA mode supports interworking between 4G and 5G networks. The NSA architecture is comprised of LTE BS (eNB), LTE evolved packet core (EPC), 5G BS (gNBs) and 5G core (5GC) network. NSA has the salient advantage of shorter implementation time as it leverages an existing 4G network with only minor modification. It can support both legacy 4G and 5G devices. Essentially, NSA mode implies multi radio access technologies (RATs) and dual connectivity for end-users [10]. Among all the NSA deployment options, 3, 4 and 7 are the most common options supporting dual connectivity and mobility robustness [11], as illustrated in Fig. 3.

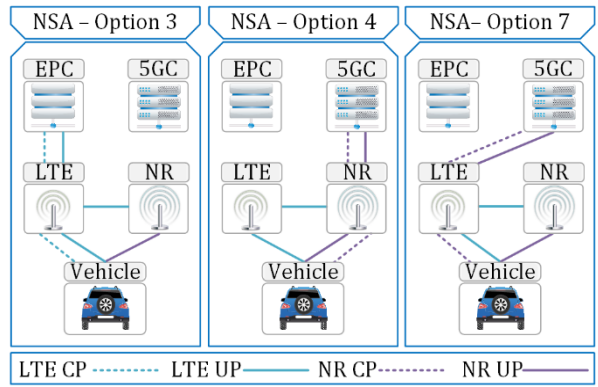

Fig. 3. NSA deployment options for 5G NR

The SA mode consists of only one technology generation, LTE or NR. The SA operation in NR is envisaged to have an entirely new end-to-end architecture and a 5GC network. In fact, gNBs, directly connected to 5GC, utilizing 5G cells for both control and user plane transfer. The SA mode is designed to enhance URLLC, whilst fulfilling the requirements of eMBB and mMTC. The key advantages of SA mode are easy deployment and improved RAT and architecture performance. That said, it requires the 5G RAT to be rebuilt and a cloud native 5G core to fully realize all the potential benefits of a 5G network. In addition, the SA mode facilitates a wider range of new use cases and supports advanced NS functions.

### D. Security Aspects of 5G NR-V2X

5G inherits the basic security mechanisms in 4G. Accordingly, NR-V2X will utilize equivalent functionalities as in the LTE-V2X. However, due to fundamental changes in 5G architecture and its end-to-end communication, new mechanisms need to be adopted. Basically, the required functionalities and security enhancement for 5G networks are largely dependent on deployment strategy. The scope of security enhancements in NSA is limited, as it is dependent on the underlying 4G deployment and requires identification of 5G functions which match to 4G components. In contrast, security enhancement with SA deployment will allow the network to support more security features to tackle potential security challenges. Next, we briefly discuss the key enhancement of NR-V2X security.

From an architectural perspective, NR-V2X should ensure security of users, vehicles, end-to-end communication entities, functions, and interfaces [12]. This can be achieved with new 5G core entities, new network functions, stronger authentication and authorization schemes between vehicles, vehicle to RAN, and vehicle to core. The security anchor function (SEAF), defined in Rel. 15, is a new function which is used to enhance security at network level and to provide flexible authentication and authorization schemes. SEAF can provide more flexible deployment of access and mobility management function (AMF) and session management function (SMF) entities. With this feature, device access authentication is separated from data session setup and management, which provides secure mobility and authorized access to V2X services for vehicles and users.

The general principle of user equipment (UE) authorization is similar to that in LTE systems. The only difference is that authorization in 5G is provided by the policy control function (PCF). Autonomous driving demands a real-time and reliable authentication process while keeping the overhead introduced by security protocols as low as possible. In terms of privacy protection, the major concern is related to encryption schemes for concealing the subscriber permanent identifier (SUPI) to protect user data leakage through initial messages. In 5G, subscriber/device privacy is provided by SUPI which is a major change from LTE with international mobile subscriber identity (IMSI). While IMSI is typically transmitted in plain text over the air, SUPI travels in ciphertext over the radio link to be protected against spoofing and tracking. Moreover, 5G enhances authentication by exploiting extensible authentication protocol (EAP) and supporting EAP authentication and key agreement (EAP-AKA) in order to separate authentication and authorization procedure in a flexible manner. The second major issue is related to user data privacy over PC5, as vehicles may need to share private information (e.g., user identity, vehicle's location). While restrictions are required with regard to the sharing of private data, some of them may need to be accessible to trusted authorities (e.g., police, rescue team) to detect malicious attackers or to ensure timely handling of emergencies such as accidents.

As far as NS is concerned, services instantiated as a NS may have different security requirements. The access to a NS should be granted only to authenticated subscribers. Security protocol should ensure communication integrity, confidentiality, and authorization. 5G introduces the concepts of slice isolation, robust slice access, and slice security management to ensure that an attack mounted on one slice does not increase the risk of attack on another slice.

In 5G NR, every gNB is logically split into central unit (CU) and distributed unit (DU). These modules interact via a secure interface. The security provided with this interface can prevent an attacker from breaching the operator's network, even in the case of successful access to the radio module.

*E.    Precise positioning*

Satellite-based positioning systems are unable to provide sufficiently accurate positioning needed for autonomous driving. LTE-V2X has been exploiting several radio signal-based mechanisms to improve the positioning accuracy, namely: downlink-based observed time difference of arrival (OTDOA), uplink time difference of arrival (UTDOA), and enhanced cell ID (E-CID). NR-V2X combines the existing positioning technologies with new positioning methods such as multicell round trip time (Multi-RTT), uplink angle of arrival (UL-AoA), downlink angle of departure (DL-AoD), and time of arrival (TOA) triangulation to provide more precise vehicle positioning [13]. Moreover, NR-V2X can also use real-time kinematic (RTK) positioning, which is an accurate satellite-based relative positioning measurement technique, to provide a centimetre-level positioning accuracy in some outdoor scenarios. By using wider bandwidth, flexible massive antenna systems, and beamforming NR-V2X will provide more precise timing and accurate measurement of equivalent signal techniques in LTE-V2X. Note that no single approach may be able to reliably provide the positioning accuracy required for autonomous driving in all environmental conditions. Hence, hybrid solutions that optimally combine NR advanced positioning techniques with multitude of embedded sensors into next generation vehicles, are the most promising approaches to achieve vehicle positioning accuracy for autonomous driving.

Table I summarizes a comparison between NR-V2X and LTE-V2X in respect to the aforementioned features.

TABLE I Comparison between LTE-V2X and NR-V2X.

| Features | LTE-V2X | NR-V2X |
| --- | --- | --- |
| Subcarrier Spacing | 15 kHz | 15,30,60,120, 240 kHz |
| Carrier Aggregation | Up to 32 | Up to 16 |
| Channel Bandwidth | 20 MHz | 400 MHz |
| Latency | < 10 ms | < 1 ms |
| Reliability | 95-99% | 99.9-99.999% |
| Channel coding | Turbo | LDPC, Polar |
| Network Slicing | No | Yes |
| Modulation | 64-QAM | 256-QAM |
| Communication Type | Broadcast only | Broadcast, Multicast, Unicast |
| Retransmission | Blind | PSFCH |
| Security and Privacy | Basic | Advanced |
| Positioning Accuracy | > 1 m | 0.1 m |

IV.  APPLICATION OF MACHINE LEARNING FOR NR-V2X

Rapidly varying vehicular environments due to vehicle mobility, frequent changes of network topology and wireless channel, as well as stringent requirements of URLLC, increase the system design complexity for a end-to-end V2X network. In such a dynamic environment, machine learning (ML) can be an effective tool to address operational challenges compared to traditional network management approaches which are more suitable for relatively low mobility scenarios. As aforementioned, vehicles are envisioned to be equipped with many on-board advanced sensors which will generate a high volume of data. In this regard, ML can efficiently analyse large volumes of data, find unique patterns and underlying structures, and finally make proper decisions by adapting to changes and uncertainties in the environment. In addition, ML can be implemented in a distributed manner to manage network issues for reduced complexity and signalling overhead as compared to a centralized approach. Thus, ML is applicable to various operational aspects of vehicular networks by using vehicle kinetics (e.g., speed, direction, acceleration), road conditions, traffic flow, wireless environments for adaptive data-driven decisions.

From the PHY perspective, in high-mobility channels, synchronization and channel estimation are challenging tasks for V2X communication system design. The V2X system may experience frequent loss of synchronization and has to deal with short-lived channel state information (CSI) estimates due to very short channel coherence times. In addition, the use of mmWave band requires fast and efficient beam tracking and switching to establish and maintain reliable links in rapidly changing environments. Here, ML can be useful in learning, tracking and predicting relevant information (i.e., the synchronization points and CSI in highly volatile channels, and beamforming directions) by exploiting historical information (i.e., user location, received power, previous beam settings, context information covering network status, and so on.).

ML may also help improve multiple UE grouping which involves cross-layer operation between PHY and medium access control (MAC). For example, should V2X use orthogonal multiple access (OMA) or non-orthogonal multiple access (NOMA) in a specific vehicular channel? Although 3GPP has decided to leave NOMA study items to beyond 5G, it is anticipated that NOMA will play an important role in autonomous driving. A major advantage of NOMA is that it serves multiple users on the same time/frequency resources. ML can help to switch between OMA and NOMA based on the requirements of each specific V2X use case.

High vehicle mobility in vehicular networks also makes the design of efficient radio resource management (RRM) mechanisms an extremely challenging problem. The conventional approach for RRM is to adopt optimization. However, due to the highly dynamic vehicular environment, traditional optimization approaches may not be feasible. For instance, a small change in the vehicular environment may require a re-run of optimization, leading to prohibitively high overhead and inefficiency. In addition, to accommodate different requirements of V2X services, multi-objective optimization could be complicated and time-consuming. ML-based approaches may be more efficient in such a scenario for a number of resource allocation problems including channel and power allocation, user association and handoff. Considering NR Sidelink transmission, vehicles are expected to reach a more sophisticated level of coordinated driving through intent sharing. In this regard, ML based transmission mode selection and resource allocation would be of interest in view of the stringent requirements in V2V communication. While resource allocation algorithms for D2D have been mostly developed in a centralized manner, the centralized control will incur huge overhead to obtain the global network information, possibly leading to bottleneck. In contrast, a decentralized ML based approach has the potential to allow every vehicle to learn the optimal resource allocation strategy from its observations. Thus, a multi-agent learning approach where each vehicle can learn and cooperate is highly desired.

Vehicle trajectory prediction has been receiving increasing interest to support driver's safe driving features such as collision avoidance and road hazard warning. While a motion model is learned based on previously observed trajectories, vehicles' future locations may be predicted via ML with the observed mobility traces and movement patterns. Additionally, unexpected factors (but affecting the trajectories) such as drivers' intention, traffic patterns, and road structures, may also be implicitly learned from historical data. Such vehicle trajectory prediction can also be helpful for handoff control, link scheduling, and routing. For instance, the most promising relay node can be selected for message forwarding and seamless handoff between V2V and V2I in an effective routing scheme using vehicle trajectory prediction.

While ML is expected to take an important role to lead data-driven intelligence and edge- and UE-based intelligence (beyond network-side intelligence), there are also challenges in adopting ML in vehicular networks. Firstly, ML may produce undesired results. While minor errors could lead to huge impacts for safety-sensitive services, significant efforts need to be made to improve the robustness and security of an ML-based approach [14]. Additionally, the on-board computational resources in each vehicle may be limited. Due to stringent end-to-end delay constraints, the use of cloud-based computing resources may not be feasible. In such cases, advanced ML techniques are needed for vehicles with limited computing capability. Thus, techniques such as model reduction or compression should be considered in designing ML-based approaches to alleviate the computation resource limitation without degradation in the performance of V2X communication.

V. 5G NR-V2X USE CASES IN COOPERATIVE AND AUTONOMOUS DRIVING

The success of 5G NR, in practice, is largely related to the question of how well 5G NR can fulfill the requirements of designated services and advanced use cases. One of the main objectives for the NR-V2X standard is to support use cases with stringent requirements of ultra-high reliability, ultra-low latency, very accurate positioning and high throughput, which may not be achieved by LTE-V2X. NR-V2X is not intended to replace LTE-V2X services but to complement them with advanced services. While LTE-V2X targets the basic safety services, NR-V2X can be used for advanced safety services as well as cooperative and connected autonomous driving. The following use cases [15] are among the target services that may be supported by NR-V2X:

- Trajectory sharing and coordinated driving: Intention/trajectory of each vehicle will be shared to enable fast, yet safe maneuvers by knowing the planned movements of surrounding vehicles. Exchange of intention and sensor data will ensure more predictable, coordinated autonomous driving, as they know the intended movements of other vehicles.
- Vehicle platooning: This is an application of cooperative driving which refers to a group of vehicles, traveling together in the same direction and at short inter-vehicle distances. To dynamically form and maintain platoon operations, all the vehicles need to receive periodic data (i.e., direction, speed and intentions) from the leading vehicle.
- Extended sensors sharing: Enables the exchange of raw or processed data gathered through local sensors or live video images among vehicles, roadside units, devices of pedestrian and V2X application servers. The vehicles can increase the perception of their environment beyond what their own sensors can detect and have a broader and more holistic view of the local situation.
- Remote driving: A remote driver or a cloud-based V2X application take control of the vehicle. Examples of remote driving/teleoperated applications are for incapacitated persons, public transportation, remote parking, logistics, or driving vehicles in dangerous environments (e.g., Mines).

The end-to-end latency and reliability requirements for the aforementioned use cases are presented in Fig. 4. As can be seen, three zones are identified. In LTE-V2X zone services which require less than 90% reliability and latency between 10-100 ms can be supported. In the second zone, service which requires 99% reliability and latency between 5-10 ms may be supported by LTE-V2X but surely are supported by NR-V2X. The services in the NR-V2X zone, which require less than 5 ms latency and above 99% reliability are only supported by NR-V2X.

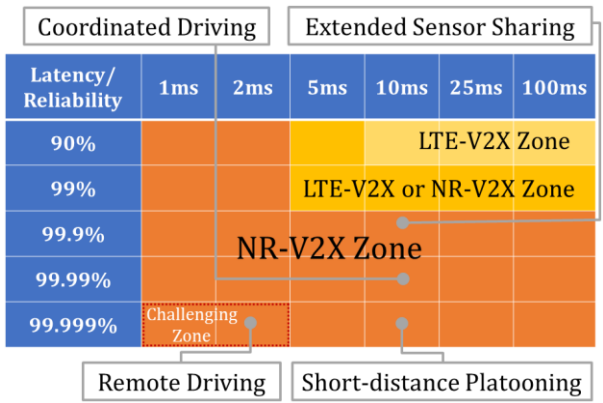

Fig. 4. End-to-end latency and reliability requirements for advanced NR-V2X use cases.

## VI. CONCLUSIONS

In this paper, we have presented the design considerations, technology components, functionalities, and key features of NR-V2X towards connected and cooperative autonomous driving. We have discussed how NR-V2X is designed and configured to fulfill a number of stringent QoS requirements associated with autonomous driving in terms of throughput, latency, reliability, security, and positioning. We have also shown that ML can be exploited to significantly improve the performances of V2X communications. It is believed that 5G NR will be a transformative technology for a highly connected and cooperative vehicular world.